\documentclass[reqno,12pt]{amsart}
\usepackage{hyperref}
\usepackage{amssymb,amsmath,amsthm}
\usepackage{a4wide}
\usepackage{amscd}
\usepackage{amsfonts}
\usepackage{amssymb}
\usepackage{latexsym}
\usepackage{color}
\usepackage{esint}
\usepackage{mathrsfs}
\newtheorem{theorem}{Theorem}

\newtheorem{lemma}[theorem]{Lemma}
\newtheorem{proposition}[theorem]{Proposition}
\newtheorem{remark}[theorem]{Remark}
\let\a=\alpha
\let\e=\varepsilon

\let\pt=\partial

\let\O=\Omega
\let\G=\Gamma
\let\o=\omega

\let\l=\lambda

\let\b=\beta

\newcommand{\sm}{\sqrt{\mu}}
\newcommand{\R}{\mathbb{R}}

\renewcommand{\P}{\mathbf{P}}

\newcommand{\ip}{(\mathbf{I}-\mathbf{P})}

\newcommand{\be}{\begin{equation}}
\newcommand{\bm}{\begin{multline}}
\newcommand{\ee}{\end{equation}}

\numberwithin{equation}{section}
\numberwithin{theorem}{section}
\date{\today}
\title[Diffusive  limit for  a Boltzmann-like equation]{Diffusive  limit for  a Boltzmann-like equation with non-conserved momentum}
\author{R. Esposito}
\thanks{(R. E.) International Research Center M\&MOCS, Università dell'Aquila, Italy}
\author{P. L. Garrido}
\thanks{(P.G.)Instituto Carlos I de Fisica Teorica y Computacional.
Universidad de Granada. E-18071 Granada, Spain}
\author{J. L. Lebowitz}
\thanks{(J. L.)Departments. of Mathematics and Physics, Rutgers University,
 USA}
\author{R. Marra}
\thanks{(R. M.) Dipartimento di Fisica and Unit\`a INFN, Universit\`a di Roma
Tor Vergata, 00133 Roma, Italy}
\begin{document}

\begin{abstract}

We consider a  kinetic model whose evolution is described by a Boltzmann-like equation for  the one-particle  phase space distribution $f(x,v,t)$. There are hard-sphere collisions between the particles as well as collisions with randomly fixed scatterers. As a result, this evolution does not conserve momentum but  only mass and energy. We prove that  the diffusively rescaled  $f^\e(x,v,t)=f(\e^{-1}x,v,\e^{-2}t)$, as $\e\to 0$ tends to a Maxwellian $M_{\rho, 0, T}=\frac{\rho}{(2\pi T)^{3/2}}\exp[{-\frac{|v|^2}{2T}}]$, where $\rho$ and $T$ are solutions of coupled diffusion equations and estimate the error in $L^2_{x,v}$.

\end{abstract}
\maketitle

\section{Introduction and results}

We study a kinetic model investigated by Garrido and Lebowitz in \cite{GL} in which  only  the mass and the energy are conserved by the evolution but not the momentum. This models the flow of a gas (or fluid)  in a porous medium. It can also be seen as the Grad-Boltzmann limit of a hard sphere system  elastically scattered by randomly distributed obstacles.   It thus serves as a simplified example for the derivation of macroscopic equations from mesoscopic kinetic ones: the number of conserved quantities is reduced from five to two. There are at present  no rigorous derivations of hydrodynamic equations in the diffusive limit when there are five conserved quantities and density and temperature are space-time dependent to the lowest order. Here we extend the heuristic analysis of this system in two dimensions described in \cite{GL} and give a fully rigorous derivation of the appropriate coupled diffusion equations.

The model is defined in the following way:

Let $\O$ be the three dimensional unit torus. The kinetic equation on $\O\times \R^3_v$ is
\be \pt_t F +v\cdot \nabla F=  Q(F) ,\label{main0}\ee
with $Q(F)=Q_B(F,F)+\alpha Q_d(F)$, $\alpha \ge 0$, 
where $Q_B(F,H)$ is the symmetrized Boltzmann collision operator for hard spheres \cite{CIP}
 defined as 
\begin{multline}
Q_B(F,H)(v):=\\\frac 1 2\int_{\mathbb{R}^{3}}\mathrm{d} w\int_{{\mathbb{S}^{2}} }\mathrm{d}\o B(v-w,\o 
)[F(v^{\prime })H(w^{\prime })+F(w^{\prime })H(v^{\prime })-F(v)H(w)-F(w)H(v)]   
\label{QB}\end{multline}
with $v^{\prime } = v-[(v-w)\cdot \o ]\o , \ w^{\prime }={v} + [(v-w)\cdot 
\o ]\o {,}$ and $B(V,\o )=|V\cdot \o |$ is \textit{the hard spheres cross section} 
and  $Q_d(F)$ models elastic collisions with randomly distributed infinite mass scatterers at rest. $Q_d(F)$ is a linear operator conserving only mass and energy, not momentum, so that 
\be \int_{\R^3_v} dv Q_d(F)=0,\quad  \int_{\R^3_v} dv |v|^2Q_d(F)=0,\label{massencons}\ee
for any $F$, but 
\be \int_{\R^3_v} dv v Q_d(F)\ne 0,\label{nomomcons}\ee
for some $F$. We also require
 that the corresponding entropy dissipation is negative
\be D_d(F)=\int dx\int dv Q_d(F)\log F \le 0 \label{dissd}.\ee
Letting $M_{\rho,u,T}$ be the local Maxwellian with $\rho$, $u$ and $T$ possibly depending on space and time 
\be M_{\rho,u,T}=\frac {\rho}{(2\pi T)^{3/2}} e^{-\frac{(v-u)^2}{2T}}.\label{maxw}\ee We set
\be \eta(\rho, u,T):=-D_d( M_{\rho,u,T})\ge 0,\quad \text{ with}\quad \eta(\rho,u, T)=0\quad \text{iff } u=0.\label{dissd1}\ee

 We can model the non momentum conserving collisions with the background  by various choices of $Q_d(F)$ \cite{GL}.
 
We prefer here, for simplicity of presentation, to consider the operator
\be Q_d(F) = \int_{{\mathbb{S}^{2}} }d\o[F(v-2(v\cdot\o) \o)- F(v)]|v\cdot \o|.\label{qd}\ee
The results in this paper apply to all choices in \cite{GL}.
Note that, since $|v-2\o(\o\cdot v)|=|v|$,  $Q_d(F)=0$ if $F$ depends on $v$ only through $|v|$.
By the Boltzmann $H$ theorem, $D_B(F)=\int_{\O}dx\int_{\R^3_v}dv \log F Q_B(F,F)\le 0$ and vanishes if and only if $F=M_{\rho,u,T}$. Moreover, if $\a>0$, we have also
\be Q_B(F,F)+\a Q_d(F)=0 \quad \text{iff } F=M_{\rho,0,T}.\label{zero}\ee
In fact, if $Q_B(F,F)+\a Q_d(F)=0$, multiplying  by $\log F$ and integrating, we obtain
\[ D_B(F)+\a D_d(F)=0.\]
But both are non positive, so we must have $D_B(F)=0$ and $D_d(F)=0$. The first implies $F= M_{\rho,u,T}$. By the second of \eqref{dissd1} then $u=0$ and we get the conclusion.

From now on we assume $\a>0$.

To look at the behavior of the solution on the diffusive space-time scale  \cite{DEL} we consider the equation for $F^\e(x,v,t)=F(\e^{-1}x,v,\e^{-2}t)$. $F^\e$ so defined satisfies the equation
\be \pt_t F^\e +\e^{-1}v\cdot \nabla F^\e= \e^{-2} Q(F^\e) ,\label{main}\ee
and we seek for its  solution in the form
\be F^\e=\mu +\sum_{i=1}^3\e^i F_i+\e^{5/2} \sqrt\mu f,\label{slit}\ee
where $\mu=M_{\rho,0,T}$.

Note that, by total mass and total energy conservation, there is no loss of generality in assuming 
\be\int_{\O\times\R^3_v} dx dv F^\e = \int_{\O\times\R^3_v} dx dv \mu, \quad \int_{\O\times\R^3_v} dx dv |v|^2F^\e = \int_{\O\times\R^3_v} dx dv |v|^2\mu.\label{normal}\ee

We prove that, as $\e\to 0$,  $F^\e(x,v,t)$ tends to the Maxwellian $M_{\rho, 0, T}$ where $\rho$ and $T$ are solutions of the following set of two coupled diffusion equations for the density $\rho$ and the temperature :
\be\begin{cases}\partial_t\rho=\displaystyle{\nabla\cdot\Big[H\frac{\nabla \rho}{\rho}\Big]+\nabla\cdot\Big[H'\frac{\nabla T}{T}\Big]}=0\\
\displaystyle{\frac 3 2}\rho\partial_tT =\displaystyle{\nabla\cdot\Big[TH'\frac{\nabla \rho}\rho\Big]+\nabla\cdot\Big[TH'_1\frac{\nabla T}T\Big]}=0\end{cases}\ee
where $H,H',H_1$ are transport coefficients whose expressions   are (independent of the index $i$)
\be H=\int dv  v_i\mathcal{L}^{-1}(\mu v_i); \quad H'=\int dv  v_i\Big(\frac{|v|^2}{2T}-\frac{3}{2}\Big)\mathcal{L}^{-1}(\mu v_i); \ee
\be H'_{1}=\int dv  v_i\Big(\frac{|v|^2}{2T}-\frac{3}{2}\Big)\mathcal{L}^{-1}\Big(v_i\Big(\frac{|v|^2}{2T}-\frac{3}{2}\Big)\mu\Big)\ee
Here 
\be\mathcal{L}F= \mathcal{L}_B F-\a Q_d(F),\label{calL}\ee
where
$$\mathcal{L}_B F=-2Q_B(\mu, F),$$
We remark that the transport coefficients $H$  and $H'$ diverge as $\a\to 0$ because $v_i\mu$ are in the null space of $\mathcal{L}_B$, and $\mathcal{L}_B^{-1}$ is not well defined on the function $\mu v_i$. 
 
Moreover, we determine also $F_1$ as 
\be F_1= -\mathcal{L}^{-1}[v\cdot \nabla\mu] + \mu\left[\frac{\rho_1}{\rho} + \frac{|v|^2-3T}{2\rho T^2}T_1\right].\label{F1bis0}\ee
where $\rho_1$, $T_1$ are solutions of  linear diffusion equations such that
\be\int_{\O} dx\rho_1 = 0, \quad \int_{\O} dx T_1 = 0.\label{norm1}\ee
$F_i$, for $i>1$, will be specified later.

The main result of this paper is the following 
\begin{theorem}\label{mainth} 
Let $t>0$ be fixed and assume that the solution $(\rho, T)$ to \eqref{diffeq} and \eqref{transp} have positive lower bounds and there is $C(\rho, T)\ll1$ such that, for $0\le k,\ell\le 4$ with $1\le k+\ell\le 4$, 
\be\sup_{0\le s\le t}\sum_{k,\ell}(\|\nabla^k\pt_t^\ell \rho(s)\|_2+\|\nabla^k \pt_t^\ell T(s)\|_2)<C(\rho,T).\label{CrhoT}    \ee Assume also that the initial value of $F$ is positive and satisfies \eqref{initial} below. Then, if $\e\ll1$,  \eqref{main} has a positive solution $F$ such that
$$ \|\mu^{-1/2}(F-\mu)\|_2\le C\e.$$
\end{theorem}
Here, for $p\ge 1$, the $\|\,\cdot\,\|_p$-norm is defined as
\be\| f\|_p=\Big[\int dx dv |f(x,v,t)|^p\Big]^{1/p};\quad \| f\|_\infty={\rm ess}\sup_{x,v}  |f(x,v,t)|
.\label{norms}\ee

\medskip

\section{Strategy of the proof}
The proof of  theorem \eqref{mainth} will be given in Section \ref{proofs} and here we only present the main ideas of the proof. 

Once the Maxwellian $\mu$ and the terms of the expansion $F_i$, $i=1,\dots,3$ in \eqref{slit} are computed, the main technical problem is to obtain bounds uniform in $\e$ for the remainder $f$, which solves a non linear problem.  To deal with the non linearity we use an iterative procedure based on two steps. The first step is to study the linear problem obtained by pretending that the non linear term is computed using the solution of the previous step of the iteration. The aim is to bound the $L^2$-norm of the solution to the linear problem (see Proposition \ref{propl2}). The novelty with respect to previous work using this ideas, e.g. \cite{EGKM1}, is the fact that the Maxwellian $\mu$ depends on $x,t$ through $\rho$ and $T$. This produces a  term singular in $\e$ in the inequality for $\|f\|_2^2$  which has to be dealt with. The most dangerous part of this term, depending on $(\P f)^2$ (here $\P$ is the projector on the space spanned by the conserved quantities) vanishes thanks to the fact that the Maxwellian $\mu$ has mean velocity $u=0$. 

The other terms which  have to be dealt with contain a polynomial of degree three in $v$ which gives troubles for large velocities. To this end, following \cite{GJJ}, we introduce a global Maxwellian $\mu_T$ with temperature given by the $\min T(x,t)$ assumed strictly positive and bound the high velocity tail of $f$ in terms of the $L^\infty$ norm of $h\sim f \sm /\sm_T$. Then $h$ is bounded in $L^\infty$ by Proposition \ref{proph}.
The presence of $\|h\|_\infty$ in the energy inequality is a serious obstacle to obtaining a global in time statement. Theorem \ref{mainth} is in fact established for arbitrary $t>0$, but with constants depending on $t$. 

Once the linear problem is solved, we need to get bounds on the non linear term. Here we have another novelty with respect to \cite{EGKM1}. In that paper the non linear term is bounded in terms of $L^p([0,t], \O, \R^3_v)$ norms of $f$ and its time derivative $f_t$, with $p=3$ and $6$. Here we cannot use this method because the equation for $f_t$ involves a term which is too singular in $\e$. Therefore we can only use $L^2$ and $L^\infty$ norms. But the singularity $\e^{-3/2}$ of the $L^\infty$ norm of $h$ (see Proposition \ref{proph}) has to be controlled  by a  sufficiently high power of $\e$ in front of the non linear term and hence we need to look for a remainder in \eqref{slit} of order $\e^{5/2}$ while in \cite{EGKM1} $\e$ was sufficient. We   remark that, as a consequence, in the present case we need to assume some regularity properties of the limiting solution, while in \cite{EGKM1}  the convergence is proved without such assumption.

\begin{remark} We conclude this section by observing that if the random collisions operator $Q_d$ is absent ($\a=0$), the problem of deriving in a rigorous way hydrodynamic equations in the diffusive limit, with non homogeneous density and temperature  at time zero, is completely open. A formal expansion shows (see e.g. \cite{DEL,B,S}) that the limiting equations are different from the Navier-Stokes equations. In stationary non homogeneous situations there are few results, see for example \cite{AEMN}.\end{remark}

\section{The expansion}
We  start presenting the expansion. Following a strategy similar to \cite{EGKM1}, \cite{EGM} (where only the first two terms of the expansion are considered) we look for a solution of the form  \eqref{slit},\eqref{normal}.
By substituting \eqref{slit}    into equation \eqref{main} we get
\begin{eqnarray*}
&-&\e^{-2}[Q_B(\mu,\mu)+\a Q_d (\mu)]+\\&&
\e^{-1}[v\cdot \nabla \mu+\mathcal{L} F_1]+\\&&
\e^0[\pt_t \mu +  v\cdot \nabla F_1+\mathcal{L} F_2-Q_B(F_1,F_1)]+\\&&
\e^1[\pt_t F_1+v\cdot\nabla F_2+\mathcal{L}F_3-2Q_B(F_1,F_2)]+\\&&
\e^2[\pt_t F_2+v\cdot \nabla F_3-2Q_B(F_1,F_3)]-Q_B(F_2,F_2)]+\\&&
\e^3[\pt_t F_3- 2Q_B(F_2,F_3)]+\e^4Q(F_3,F_3)]+
\\&&\e^{5/2}\Big[\pt_t (\sm f)+\e^{-1} v\cdot \nabla (\sm f) +\e^{-2}\mathcal{L}(\sm f)\\&&-\e^{-1}\Big(2Q_B(F_1,\sm f)+2\e Q_B(F_2,\sm f)+2 \e^2 Q_B(F_3,\sm f)\Big) -\e^{1/2} Q_B(\sm f,\sm f)\Big]=0.
\end{eqnarray*}

We now examine all the terms in the equation above. The most diverging  term  $\e^{-2}[Q_B(\mu,\mu)+Q_d(\mu)]$ vanishes  because $\mu$ is a local Maxwellian with vanishing mean velocity. 

To cancel the  diverging term of order  $\e^{-1}$ we impose
\be v\cdot \nabla \mu + \mathcal{L} F_1=0.\label{F1}\ee
Since $\int dv \mathcal{L} F=0$ and $\int dv |v|^2\mathcal{L} F=0$, the previous equation has a solution under the solvability conditons 
\be\int dv  v\cdot \nabla \mu=0, \quad \int dv |v|^2 v\cdot \nabla \mu=0.\label{compat}\ee
But $v\cdot \nabla\mu$ is odd in $v$, so  conditions \eqref{compat} are satisfied and one can write the most general solution to \eqref{F1}   in the form
\be F_1= -\mathcal{L}^{-1}[v\cdot \nabla\mu] + \mu\left[\frac{\rho_1}{\rho} + \frac{|v|^2-3T}{2\rho T^2}T_1\right],\label{F1bis}\ee
where the second term is the component of $F_1$ in the null space of $\mathcal{L}$. One can choose  $\rho_1$, $T_1$ such that
\be\int_{\O} dx\rho_1 = 0, \quad \int_{\O} dx T_1 = 0.\ee

The term of order $\e$ satisfies the equation
\be \pt_t\mu+ v\cdot \nabla F_1-Q_B(F_1,F_1)+\mathcal{L} F_2=0,\label{eqF2}\ee
This equation has a solution if the following solvability conditions are satisfied
\be \int dv [\pt_t\mu+ v\cdot \nabla F_1]=0, \quad \int dv |v|^2[\pt_t\mu+ v\cdot \nabla F_1]=0.\label{compatF2}\ee

Hence
\be F_2= -\mathcal{L}^{-1}\Big[\pt_t\mu +v\cdot\nabla F_1-Q_B(F_1,F_1) \Big]+\mu\left[\frac{\rho_2}{\rho} + \frac{|v|^2-3T}{2\rho T^2}T_2\right].\label{exprF2}\ee
By using the expression of $F_1$ in \eqref{F1bis}  the  solvability conditions \eqref{compatF2} provide two diffusion equations for the density and temperature. In fact, writing \eqref{compatF2} explicitly we get
\be \int dv \left\{\pt_t\mu- v\cdot \nabla \left[\mathcal{L}^{-1}[v\cdot \nabla\mu] + \mu\left[\frac{\rho_1}{\rho} + \frac{|v|^2-3T}{2\rho T^2}T_1\right]\right]\right\}=0, \label{diff_eq1}\ee
 \be\int dv |v|^2\left\{\pt_t\mu- v\cdot \nabla \left[\mathcal{L}^{-1}[v\cdot \nabla\mu] + \mu\left[\frac{\rho_1}{\rho} + \frac{|v|^2-3T}{2\rho T^2}T_1\right]\right]\right\}=0\label{diff_eq2}\ee
 The last term in both equations vanishes because it is odd in $v$. 
Thus, the equations reduce to
\be \int dv \left\{\pt_t\mu- v\cdot \nabla \mathcal{L}^{-1}[v\cdot \nabla\mu] \right\}=0, \label{diff_eq1.1}\ee
 \be\int dv \frac{|v|^2}{2}\left\{\pt_t\mu- v\cdot \nabla \mathcal{L}^{-1}[v\cdot \nabla\mu] \right\}=0\label{diff_eq2.1}\ee
The term of order $\e$ is canceled by requiring 
\be \pt_t F_1+v\cdot\nabla F_2+\mathcal{L}F_3-2Q_B(F_1,F_2)=0\label{LF3}.\ee
We  use this to find $F_3$:
\be F_3=-\mathcal{L}^{-1}\Big[ \pt_t F_1+v\cdot \nabla F_2- Q_B(F_1,F_2)\Big]+\mu\left[\frac{\rho_3}{\rho} + \frac{|v|^2-3T}{2\rho T^2}T_3\right] .\label{eqF3}\ee
provided that 
\be \int dv (\pt_t F_1+ v\cdot \nabla F_2)=0, \quad \int dv |v|^2(\pt_t F_1+ v\cdot \nabla F_2)=0.\label{compatF3}\ee
Finally, we are left with an equation for $f$ which will be discussed later.

\medskip

Now we proceed by finding the explicit expression of the hydrodynamic equations \eqref{diff_eq1.1} and \eqref{diff_eq2.1}. 

We have:
\be\partial_t\rho-\sum_{i,j}\pt_i\big[\int dv v_i\mathcal{L}^{-1}(v_j\mu)\frac{\pt_j\rho}{\rho}\big]-\sum_{i,j}\pt_i\big[\int dv v_i\mathcal{L}^{-1}(v_j\mu (\frac{|v|^2}{2T}-\frac{3}{2}))\frac{\pt_j T}{T}\big]=0\ee
\begin{multline} \frac{3}{2}\partial_t(\rho T) -\sum_{i,j}\pt_i\big[T\int dv (\frac{|v|^2}{2T}-\frac{3}{2})v_i\mathcal{L}^{-1}(v_j\mu)\frac{\pt_j\rho}{\rho}\big]-\frac{3}{2}\sum_{i,j}\pt_i\big[T\int dv v_i\mathcal{L}^{-1}(v_j\mu )\frac{\pt_j \rho}{\rho}\big]\\-\sum_{i,j}\pt_i\big[T\int dv v_i(\frac{|v|^2}{2T}-\frac{3}{2})\mathcal{L}^{-1}(v_j\mu (\frac{|v|^2}{2T}-\frac{3}{2}))\frac{\pt_j T}{T}\big]\\
-\frac{3}{2}\sum_{i,j}\pt_i\big[T\int dv v_i\mathcal{L}^{-1}(v_j\mu (\frac{|v|^2}{2T}-\frac{3}{2}))\frac{\pt_j T}{T}\big]=0\end{multline}
 We notice  that $\mathcal{L}^{-1}(v_j\mu (\frac{|v|^2}{2T}-\frac{3}{2}))$ and $\mathcal{L}^{-1}(v_j\mu)$ are well defined since the functions $v_j\mu (\frac{|v|^2}{2T}-\frac{3}{2}$ and $(v_j\mu)$ are in the space orthogonal to the null space of $\mathcal{L}$. 
 We can use that $\mathcal{L}^{-1}$ is self-adjoint with respect to the $L^2$ scalar product with weight ${\mu}^{-1}$ to write
\be\int dv (\frac{|v|^2}{2T}-\frac{3}{2})v_j\mathcal{L}^{-1}(v_i\mu)=\int dv v_i\mathcal{L}^{-1}(v_j \mu(\frac{|v|^2}{2T}-\frac{3}{2}))\ee Define the transport coefficients:
\be H_{ij}=\int dv  v_j\mathcal{L}^{-1}(\mu v_i);; \quad H'_{ij}=\int dv  v_j(\frac{|v|^2}{2T}-\frac{3}{2})\mathcal{L}^{-1}(\mu v_i); \ee
\be H'_{1_{ij}}=\int dv  v_j(\frac{|v|^2}{2T}-\frac{3}{2})\mathcal{L}^{-1}(v_i(\frac{|v|^2}{2T}-\frac{3}{2})\mu)\ee
For isotropy reasons 
$H_{ij}=\delta_{ij} H; H'_{1ij}=\delta_{ij}H'; H'_{{ij}}=\delta_{ij}H'_1$.

Then, the previous equations take the form:
\be\partial_t\rho=\nabla\cdot \big[H\frac{\nabla\rho}{\rho}\big]+\nabla\cdot \big[H'\frac{\nabla T}{T})\big]\label{diffeq}\ee
\be \partial_t e =\nabla\cdot \big[(TH'+\frac{3}{2}TH)\frac{\nabla\rho}{\rho}\big]+\nabla\cdot\big[(TH_1'+\frac{3}{2}TH')\frac{\nabla T}{T}\big]\label{transp}\ee
where $e=\frac{3}{2}\rho T$ is the internal  energy density.

To relate the above transport coefficients to the Onsager coefficients, we introduce the fugacity $z$ (related to the chemical potential $\mu$ through $\frac{\mu}{T}=\log z$) given for the perfect gas in $3d$ by 
$\log z=\displaystyle{\log \frac{\rho}{T^{3/2}}}$ and write the equations in the form
\be\partial_t\rho=-\nabla\cdot J_\rho\label{eqrho}\ee
\be \partial_t e =-\nabla\cdot J_e\label{eqT}\ee
where
\be J_\rho=-L_{\rho\rho}\nabla{\log z}+L_{\rho e}\nabla \frac{1}{T}\ee
\be J_p=-L_{ e\rho}\nabla {\log z}+L_{ee}\nabla \frac{1}{T}\label{eqT1}\ee
so that
\be\partial_t\rho=\nabla\cdot (L_{\rho\rho}\frac{\nabla\rho}{\rho})-\nabla\cdot[(\frac{3}{2}L_{\rho\rho}-\frac{L_{\rho e}}{T}) \frac{\nabla T}{T}]\ee
\be \partial_t e=\nabla\cdot(L_{ e\rho}\frac{\nabla\rho}{\rho})-\nabla\cdot[(\frac{3}{2}L_{ e\rho}-\frac{L_{ee}}{T})\frac{\nabla T}{T}]\label{eqT2}\ee

By comparison,
$$H=-L_{\rho\rho};\quad T(H'+\frac{3}{2}H)=-L_{\rho e}=-L_{e\rho};\quad T^2(H_1'-\frac{9}{4}H)=L_{ee}$$

In \cite{GL} the form of the transport coefficients for  a different choice of the  linear Boltzmann operator   and in dimension $2$ is discussed. 

To get the regularity properties of the transport coefficients  we can for example use the method in \cite{BM}, where the eigenvalues of the linearized and the linear Boltzmann operators are studied by an expansion in spherical functions.

Plugging the expression for $F_1$ \eqref{F1bis} and the one for $F_2$ \eqref{exprF2} in the first equation of \eqref{compatF3} we get
\begin{multline} \int dv\big\{\pt_t(-\mathcal{L}^{-1}[v\cdot \nabla\mu] + \mu\left[\frac{\rho_1}{\rho} + \frac{|v|^2-3T}{2\rho T^2}T_1\right])\big\}\\
+\int dvv\cdot \nabla\big[\mathcal{L}^{-1}\big[\pt_t\mu +v\cdot\nabla F_1-Q_B(F_1,F_1) \big]\big\}.\label{exprF20}
\big]\big\}=0\end{multline}
The first  and the third term  in the first integral and the first term in the second integral do not give  contribution. We are left with
$$\pt_t \rho_1+\pt_i \int dvv_i\mathcal{L}^{-1}\big [v\cdot\nabla [\mu(\frac{\rho_1}{\rho} + \frac{|v|^2-3T}{2\rho T^2}T_1)]-Q_B(F_1,F_1)\big ] =0$$ 
By using the relation $Q_D(h,h)=-\mathcal L(\mu^{-1}h^2)$ and noting that $\int dvv_i\mathcal{L}_B^{-1}[\mathcal{L}_B\displaystyle{\frac{F_1^2}{\mu}}]=0$ by oddness, we get

$$\pt_t \rho_1+\pt_i \int \mathcal{L}^{-1}(\mu v_i)v_j\pt_j[\frac{\rho_1}{\rho} + \frac{|v|^2-3T}{2\rho T^2}T_1]+\pt_i \int \mathcal{L}^{-1}(\mu v_i)v_j[\frac{\rho_1}{\rho} + \frac{|v|^2-3T}{2\rho T^2}T_1]\frac{\pt_j\mu}{\mu}=0$$ 
  which is   a  linear diffusive non homogenous equation for $\rho_1$

By proceeding in the same way starting from the second compatibility condition, we get a linear  diffusive non homogeneous equation for $T_1$.

The compatibility conditions \eqref{compatF2}, \eqref{compatF3} thus turn out to be diffusion equations for $\rho$, $T$, $\rho_1$ and $T_1$. On the other hand $\rho_2$ and $T_2$ are determined by the orthogonality condition \eqref{orthA} discussed in Section \ref{proofs}, which are also diffusion equations; the procedure to get the equations for them  is the same and we omit it. $\rho_3$ and $T_3$ are determined by a different condition \eqref{orthA2}, also discussed in Section \ref{proofs}.

\section{Proofs}\label{proofs}
In this technical section  we construct a solution for the equation for the remainder  and prove Theorem \ref{mainth}.
\subsection{General setup}\label{gensetup}
We need some notation. We denote  
\be Lf =\mu^{-1/2}\mathcal{L}(\sm f).\ee
Since $\mu(v-2\o(\o\cdot v))=\mu(v)$, we have
\be{L}f={L}_Bf + L_d(f),\label{L}\ee where 
\be {L}_Bf=-\mu^{-\frac 1 2}\mathcal{L}_B(\sm f),\quad L_d f= -\a Q_d(f).\label{LB}\ee
Below, depending on the context, $(f,g)$ denotes the standard $L^2$ inner product in $L^2(\R^3_v)$ or in $L^2(\Omega\times \R^3_v)$. As is well known (see e.g. \cite{CIP}), the quadratic form $(f,{L}_B f)$ is non negative and strictly positive if $f$ belongs to the orthogonal complement of the null space of $\mathcal{L}_B$, which is spanned by the orthonormal functions 
\be\psi_m= \sqrt{\rho}^{-1}\sm,\quad \psi_i=\quad {\sqrt{\rho T}}^{-1}v_i {\sm},\quad i=1,\dots,3,\quad\psi_e =\sqrt{6\rho T^2}^{-1}(|v|^2-3T)\sm.\label{nullB}\ee
On the other hand, a direct check shows that there is $\l_d>0$ such that 
\be (\psi_i, L_d \psi_j)= \l_d \delta_{ij} \quad \text{for } i,j=1,\dots,3.\label{eig}\ee

Let $\mathcal {N}$ be the null space of ${L}$. By the previous observations we immediately conclude that it is the  linear subspace spanned by the normalized vectors 
\be\psi_m= \sqrt{\rho}^{-1}\sm, \quad\psi_e =\sqrt{6\rho T^2}^{-1}(|v|^2-3T)\sm.\label{null}\ee
In fact, let us consider the quadratic form
\be (f,L f) = (f, L_B f) + (f, L_d f).\label{quad}\ee
If $f\in \mathcal {N}$ then $(f,Lf)=0$. Since the two terms in \eqref{quad} are both non negative, then $(f,L_Bf)=0$ and $(f, L_d f)=0$. Thus $f\in \mathcal{N}_B$ and $(f,\psi_i)=0$ for $i=1,\dots,3$. Hence $\mathcal{N}$ is spanned by \eqref{null}.

Denote by $\P$ the projector on such a subspace and by $\ip$ the projector on the orthogonal subspace. Then
\be \P L=L \P=0.\ee

For the Boltzmann collisions  Grad proved (see \cite{Gr}), 
\be L_Bf=\nu_B f-K_B f\ee
where $K_B$ is a compact operator and $\nu_B$ satisfies the following bounds: there are positive $\tilde\nu_0$ and $\tilde\nu_1$ such that
\be 0<\tilde\nu_0\langle v\rangle\le \nu_B(x,v)\le \tilde\nu_1\langle v\rangle,\ee
with $\langle v\rangle= (1+|v|^2)^{\frac 1 2}$. The statement can be easily extended to the present  operator $L$ which can thus be decomposed as
\be Lf=\nu f-Kf\ee
where $K$ is a compact operator and $\nu$ satisfies the following bounds which follow immediately from \eqref{qd}: there are positive $\nu_0$ and $\nu_1$ such that
\be 0<\nu_0\langle v\rangle\le \nu(x,v)\le \nu_1\langle v\rangle.\ee
The following spectral inequality holds:
\be (f,L f)\ge \l\|\ip f\|_\nu^2,\label{spin}\ee
for a positive $\l$ and
\be \|f\|_\nu=\|\sqrt{\nu} f\|_2.\ee

To prove \eqref{spin}, we note, as is well known (see e.g. \cite{CIP}), that for the linear Boltzmann operator $L_B$ the following spectral inequality holds:
\be (f,L_B f)\ge \l_B\|(\mathbf{I}-\P_B) f\|_\nu^2,\label{spinB}\ee
where $\P_B f=\P f +\sum_{j=1}^3 (f,\psi_j)\psi_j$ and $\psi_j=\sqrt{\frac{\mu}{\rho T}}v_j$. By \eqref{eig} we obtain \eqref{spin} with $\lambda =\min (\lambda_B,\lambda_d)$. Note that $\l_d\sim \a$. Thus $\l_d\to 0$ as $\a\to 0$.

\medskip
The main core of the proof of Theorem \ref{mainth} is the control of the remainder $\sqrt\mu f$ which satisfies the following equation
\begin{multline}\pt_t (\sm  f) +\e^{-1} v\cdot \nabla (\sm  f)=\\-\e^{-2} \mathcal{L} (\sm  f) +\e^{-1}\mathcal{L}^1 (\sm  f) +\e^{1/2}Q_B(\sm  f,\sm  f)+\e^{-1/2}  \tilde A
,\label{eqtildef}\end{multline}
where \be\mathcal{L}^1 f=2Q_B(F_2+\e F_2+\e^2 F_3, f)\label{L1}\ee and 
\be\tilde A=[Q_B(F_2,F_2)+2Q_B(F_1,F_3)+ 2\e Q_B(F_2,F_3)+\e^2 Q(F_3,F_3)-\pt_t F_2-\e\pt_t F_3- v\cdot \nabla F_3].\label{A}\ee

For the estimate of $f$ it will be essential that the part of $\tilde A$ in the null space of $\mathcal{L}$ vanishes. For this reason we also impose that 
\be \int dv (\pt_t F_2+ v\cdot\nabla F_3)=0, \quad \int dv |v|^2(\pt_t F_2+ v\cdot\nabla F_3)=0.\label{orthA}\ee
\be \int dv \pt_t F_3=0, \quad \int dv |v|^2\pt_t F_3=0.\label{orthA2}\ee
As before, \eqref{orthA} becomes a couple on linear non homogeneous parabolic equations.

Finally, we use the freedom of choice of $\rho_3$ and $T_3$ to ensure \eqref{orthA2}. 

We have the following
\begin{proposition}\label{macrosc}
Assume $0<\rho_0(x)$, $0<T_0(x)$ with finite $L^\infty$ norms. Then, for any $\bar t>0$ the equations \eqref{diffeq} and \eqref{transp} have smooth solutions. Moreover, if \eqref{CrhoT} is verified at time $t=0$, then it stays true for $t\in [0,\bar t]$. 
The functions $F_i$ satisfy the inequalities
\be \|\mu^{-1/2}F_i\|_2+\|\mu^{-1/2}F_i\|_\infty \le C(\rho,T),\ee
for $i=1,\dots,3$.
\end{proposition}
Proposition \ref{macrosc} is a simple consequence of the parabolic regularity and we omit the proof (see e.g. \cite{Fried}).

It is convenient to write  
\be \tilde A =\sm A, \quad \G(f,g)=\sm^{-1}Q_B(\sm f,\sm g).\label{defgamma}\ee
We note that
\be \quad \P \G(f,g)=0.\label{PGamma=0}\ee
The equation for $f$ then becomes
\be\pt_t  f +\e^{-1} v\cdot \nabla  f+ \frac 1 2f[(\pt_t+\e^{-1} v\cdot \nabla) \log\mu] +\e^{-2} {L}  f =\e^{-1}{L}^1  f +\e^{\frac12}\G( f, f)+ \e^{-\frac 1 2} A,\label{eqf}\ee
where
\be L^1 f= 2 \G(f_1+\e f_2+\e^2 f_3, f),\label{defL1}\ee
and
\begin{multline} A=[\G(f_2,f_2)+2\G(f_1,f_3)-\mu^{-1/2}(\pt_t (\sm f_2)+v\cdot \nabla (\sm f_3)]\\+2\e \G(f_2,f_3)+\e^2 \G(f_3,f_3)-\mu^{-1/2}\e\pt_t (\sm f_3),\label{defA}\end{multline}
where we have set $f_i=\mu^{-1/2}F_i$.
Equation \eqref{eqf} has to be solved with initial datum $f(0)$ such that
\be F(0)=\mu(0)+ \e F_1(0)+\e^2F_2(0)+\e^3 F_3(0)+\e^{5/2} \sm(0) f(0)>0.\label{initial}\ee
It is standard to check (see e.g. \cite{EGKM1}) that it is possible to construct $f(0)$ so that \eqref{initial} is satisfied.

\medskip
It will be essential to have $\P A=0$.
To this end we have  required \eqref{orthA} and \eqref{orthA2} which imply
\be \P[\mu^{-1/2}(\pt_t (\sm f_2) +v\cdot \nabla (\sm f_3)]=0,\quad \P [\mu^{-1/2}\pt_t(\sm f_3)]=0\ee
and hence 
\be\P A =0\label{PA=0}.\ee

\medskip

We note the presence in \eqref{eqf} of the divergent term $\frac 1 2f[\e^{-1} v\cdot \nabla \log\mu]$, which is a source of extra difficulties. It is due to the use of the decomposition $F=\mu +\e F_1+\e^2 F_2+ \e^3 F_3+\e^{5/2}\sm f$ instead of $F+\e F_1+\e^2 F_2+ \e^3 F_3+\e^{5/2} R$. However the first decomposition is useful to take advantage of the spectral properties of the operator $L$ in $L^2(\R^3_v)$. Alternatively, we could use the second decomposition, but then the spectral properties we need should be sought for in $L^2(\R^3_v, \mu^{-1})$ and, when integrating by parts, the weight $\mu^{-1}$ would produce a similar term in the resulting equation for $\|\mu^{-1/2}R\|_2$. As we shall see, the spectral properties are crucial in our proof, so we need to deal with this extra term.

We will follow the approach in \cite{GJJ} and \cite{EGKM1}. Instead of repeating all the proofs in these papers, we only outline and give explicit proofs when the previous approach has to be modified to adapt to the case we study. The difference with respect to \cite{GJJ} is that the scaling in this case is diffusive instead of hyperbolic. The main difference with respect to \cite{EGKM1} is in getting  a $L^2$ bound for the linear  equation due to the presence of the third term in \eqref{ene} below. This term appears because $\mu$ depends on $x$ and $t$. 

First of all we remark that in this term there will be a contribution including  a higher power of velocity, $|v|^3 f$,which is not present in the case studied in \cite{EGKM1}. 
Following \cite{GJJ}, we introduce $L^2$ and $L^\infty$ polynomial norms to control it. 

By the assumption that $C(\rho,T)$ is sufficiently small, we see that there is $T_M>0$ such that for any $0\le s\le t$ and $x\in\O$,
\be T_M<  T(s,x)<2T_M \label{inTM}.\ee
We define the global Maxwellian
\be \mu_M = \frac{1}{(2\pi T_M)^{3/2}}\exp\Big\{\frac{-|v|^2}{2  T_M}\Big\}. \label{defMT}\ee
The inequalities \eqref{inTM} imply that  
there exist constants $c_1$ and $ c_2$ such that for some $1/2 <
\alpha< 1$ and for each $(t , x , v)$ 
\be c_1\mu_M \le \mu\le c_2\mu_M^\alpha .\label{ineqmu}\ee
We stress that the previous bounds are true under the assumption of a slowly varying $T(x,t)$. 
Furthermore, we introduce the polynomial $w_\sigma(v)=(1+|v|^2)^{\sigma}$ for the control of the cubic power of velocity. We define $h$ by the position 
\be f=((1+|v|^2)^{-\sigma}\sqrt{\frac{\mu_M}{\mu }}h
,\label{defh}\ee
and choose $ \sigma>0$ later. Note that in consequence of \eqref{ineqmu} we have
\be |f(x,t)|\le C |h(x,t)|,\label{f<h}\ee
for some $C>0$.

\subsection{The linear Problem}
We start with the linear equation  
\be\pt_t  f +\e^{-1} v\cdot \nabla  f+ \frac 1 2f[(\pt_t+\e^{-1} v\cdot \nabla) \log\mu] +\e^{-2} {L}  f =g,\label{ene}\ee
with some $g\in L^2(\Omega\times \R_v)$, such that,
\be\P g=0. \label{Pg=0}\ee
In applying the result, we shall use $g$ of the form
\be g= \e^{-1}L^1 f+\e^{\frac 1 2}\G(f,f) +\e^{-\frac 1 2} A,\label{defg}\ee
so that \eqref{Pg=0} is satisfied.

As we shall see, the use of the spectral inequality \eqref{spin} in the energy inequality provides the control of the $L^2$-norm of $\ip f$. What is missing is the control of the $L^p$-norm of $\P f$. This is achieved as in \cite{EGKM1}, which can be extended to the present setup.
We have the following
\begin{proposition}\label{prop2}
Suppose $\P g=0$. If $\e>0$ is sufficiently small, there exists
a function $G(t)$ such that, for all $0\le s\leq t$, $|G(s)|\le 
\|f(s)\|_{2}^{2}$ and a constant $C$ so that 
\begin{equation}
\int_{s}^{t}\|\P f(\tau)\|_{\nu }^{2} \le C\left[
G(t)-G(s)+\e^2\int_{s}^{t}\| \nu^{-\frac 1 2}g(\tau) \|_{2}^{2}
+ \e^{-2}\int_{s}^{t}\|\ip f(\tau)\|_{\nu }^{2}\right].\label{ass}
\end{equation}
\end{proposition}
The proof of this proposition is postponed to the next section.
Note  that in \cite{EGKM1} it is only requested that $\int dx dv g=0$, clearly implied by the stronger condition $\P g=0$.
\medskip

As we shall see, the control of the cubic in $v$ term appearing in the energy inequality requires an $L^\infty$ estimate of the function $h$ defined in \eqref{defh}. 
This will be achieved by using the proposition below.
As a consequence of  \eqref{ene}, $h$ satisfies the equation
\be\partial_t h+\frac{1}{\e}v\cdot\nabla  h +\frac{1}{\e^2}L_M h=
\tilde g,\label{eqh}
\ee

with 
\be L_M  h=\frac{w}{\sqrt\mu_M}\mathcal L (\frac{1}{w}\sqrt{\mu_M}  h),\quad \tilde g=w\sqrt{\frac{\mu}{\mu_T}}g,\ee
where $w=(1+|v|^2)^{\sigma}$.
\begin{proposition}\label{proph}
If $f$ solves \eqref{ene} with $g$ given by \eqref{defg} and $\rho$, $T$ are smooth solutions to \eqref{eqrho} and \eqref{eqT} with stricly positive lower bounds uniform in $t$, such that inequalities \eqref{ineqmu} are satisfied, then there is a constant $C$ such that, for $\e$ sufficiently small,
\be 
\sup_{0\le s\le t}\|\e^{3/2} h(s)\|_\infty\le C\Big[\e^{3/2}\|h_0\|_\infty+\sup_{0\le s\le t}\|f\|_2 +\e^{7/2}\|\frac{1}{1+|v|^2}\tilde g\|_\infty\Big].\label{esth}\ee
\end{proposition}
The proof is given in \cite{EGKM1} and \cite{GJJ}. 

\medskip

This proposition will be used within an iterative procedure where, at some step we know $\bar h$ and want to compute $h$ using \eqref{eqh}  with 
\be\tilde g={\e^{-1}}L_M^1\bar h +\e^{-1/2}\frac{w}{\sqrt\mu_M}\tilde A+\e^{1/2}\frac{1}{2}\frac{w}{\sqrt\mu}Q_B(\frac{1}{w}\sqrt{\mu_M} \bar h,\frac{1}{w}\sqrt{\mu_M} \bar h), \label{tildeg}\ee
where 
\be L _M^1\bar h=\frac{w}{\sqrt\mu_M}\mathcal L^1(\frac{1}{w}\sqrt{\mu_M} \bar h)\label{LM}\ee
We shall use the bounds \cite{GJJ}:{
\be  |L _M^1\bar h|_\infty\le \nu_M\|\bar h\|_\infty\|\frac{w}{\sqrt\mu_M}\sum_{i=1}^3\e^{i-1} F^i\|_\infty\le \nu_M C(\rho,T)\|\bar h\|_\infty,\ee
\be\Big|\frac{w}{\sqrt\mu_M}Q_B(\frac{1}{w}\sqrt{\mu_M} \bar h,\frac{1}{w}\sqrt{\mu_M} \bar h)\Big|\le \nu(\mu)\|\bar h\|_\infty^2.\ee}

\medskip
To get an $L^2$ bound on the linear equation we multiply it by $f$ and integrate in $x$ and $v$ to obtain:
\be\frac 1 2 \frac{d}{dt}\|f\|_2^2+\int dxdv f \partial_t\log\mu+\frac 12\e^{-1}\int dxdv f^2v\cdot\nabla \log \mu+\e^{-2}(f,Lf)=(f,g).\label{ener0}\ee 

\medskip

\begin{proposition}\label{propl2} 
Suppose that $C(\rho,T)$ is sufficiently small. If $f$ solves \eqref{ene} 
with $g$ satisfying  \eqref{Pg=0} 
and $\rho$, $T$ are smooth solutions to \eqref{eqrho} and \eqref{eqT} with stricly positive lower bounds uniform in $t$, then, fixed $t>0$, there is a constant $C$ such that, for $\e$ sufficiently small,
\begin{multline}\|f(t)\|_2^2+\e^{-2}\int_0^{ t} ds\|\ip f(s)\|_\nu^2\\\le \e^2\int_0^{ t} ds\|\nu^{-1/2} g(s)\|_2^2+C_\kappa C(\rho,T) \e^4\int_0^{ t}\|h(s)\|^2_\infty+2\|f(0)\|_2^2.\label{ener040}\end{multline}
\end{proposition}

\medskip
\begin{proof}
In equation \eqref{ener0} the essential difference w.r.t. \cite{EGKM1} are the second and third terms. 
To bound the third term we split $f^2= [\P f]^2+ [\ip f]^2 +2\P f \ip f$ and compute the contributions separately. The most singular (in $\e$) one is due to $|\P f|^2$, whose size in $L^2$ is $\e^{-2}$-times larger than $\|\ip f\|_2^2$ by Proposition \ref{prop2}. 
Fortunately, we have
\be\int dxdv (\P f)^2v\cdot\nabla \log \mu= \int dxdv [a+ c(|v|^2-3T)]^2\mu v\cdot\nabla \log \mu=0,\label{bad}\ee
because $v\cdot\nabla \log \mu$ is odd in $v$, while $a+ c(|v|^2-3T)$ is even in $v$. Thus the largest term vanishes.
As for the term $\e^{-1}\int dxdv \P f \ip f v\cdot\nabla \log \mu$, we note that
$$ \|[v\cdot\nabla \log \mu]\nu^{-1/2} \P f \|_2\le C (\rho,T)\|\P f\|_2$$
because the factor $\sm$ in $\P f$ controls the  polynomial $\nu^{-1/2}v\cdot\nabla \log \mu$.
Thus
\begin{multline}\Big| \e^{-1}\int dxdv \nu^{-1/2}v\cdot\nabla \log \mu\P f\nu^{1/2}\ip f \Big|\le C\e^{-1}C(\rho,T)  \|\P f\|_2\|\ip f\|_\nu\le\\
\frac 1 2C(\rho,T)\|\P f\|_2^2 +\frac 1 2\e^{-2}C(\rho,T)\|\ip f\|_\nu^2\end{multline}
and the first term is controlled using Proposition \ref{prop2}.

Now, as in \cite{GJJ} we introduce a cut-off on the velocity ${\kappa}{\e^{-a}}$ for some $a>0$ to be chosen, and estimate separately the term with low and high velocity.

We bound $\frac 1 2\e^{-1}\int dxdv (\ip f)^2 v\cdot\nabla \log \mu$.  We have
$$\Big|\e^{-1}\int dxdv (\ip f)^2 v\cdot\nabla \log \mu\Big|\le  \e^{-1}\int_{v\le \kappa\e^{-a}}+ \e^{-1}\int_{v\ge \kappa\e^{-a}}
.$$
The  first term is bounded as 

\begin{multline} \e^{-1}C(\rho,T)\|[(1+v^2)^{3/4}\nu^{-1/2}I_{v\le \frac{k}{\e^a}}\ip f\nu^{1/2}]^2\le\\
C\e^{-1}\left(\frac{\kappa}{\e^a}\right)^2C(\rho,T )\|\ip f\|_\nu^2.
\end{multline}
The high velocity part is bounded as
$$  \e^{-1}C(\rho,T )\|(1+v^2)^{3/2}\nu^{-1/2}f I_{v\ge \frac{\kappa}{\e^a}}\|_\infty\|\ip f\|_\nu\le C_\kappa C(\rho,T)\e\|h\|_\infty\|\ip f\|_{\nu},$$
where $C_k= {\frac 1 {(\e^{2a}+\kappa^2)^2}}$. Moreover
we have used  that $\mu_M\le C\mu$ and $|(1+v^2)^{3/2}\nu^{-/2}f|\le |(1+v^2)^{-2}h|$ for $\sigma>\frac 5 4+2$ and $a=\frac 1 2$. Now, 
\begin{multline}C_\kappa C(\rho,T)\e\|h\|_\infty\|\ip f\|_2\le C_\kappa C(\rho,T) \e[\delta\|h\|_\infty^2+{\frac 1 {4\delta}}\|\ip f\|_\nu^2]\le\\
C_\kappa C(\rho,T) \e^4\|h\|_\infty^2+\frac 1 4C_\kappa C(\rho,T)\e^{-2}\|\ip f\|_\nu^2],\end{multline}
by choosing $\delta=\e^3$.

The term
$$\int_0^t ds \frac 12\int dxdv f^2\pt_t\log\mu$$ also contains a contribution involving $|\P f|^2$ which in this case is not zero. We have
\be \int_0^t ds \frac 12\int dxdv (\P f)^2\pt_t\log\mu\le C(\rho,T)\int_0^t ds (\|a(s)\|^2_2+\|c(s)\|^2_2), \ee 
  so that  Proposition \ref{prop2} can be used. The other terms can be controlled as before. Note that there is no $\e^{-1}$ factor.

Next, since $\P g=0$, we have the bound
\be |(f,g)|\le \gamma\e^{-2} \|\ip f\|_{\nu}^2 +\frac 1{4\gamma} \e^2\| \nu^{-1/2}g\|_2^2\label{fg}
\ee 
Summarizing, by using \eqref{spin} we have
\begin{multline}\frac 1 2 \frac{d}{dt}\|f\|_2^2+\e^{-2}\|\ip f\|_\nu^2[\lambda-\gamma-C(\rho,T)(C\kappa^2+C_\kappa)]\le\\\frac{\e^2}{4\gamma}\|\nu^{-1/2} g\|_2^2 + CC(\rho,T)\|\P f\|_2^2 + C_\kappa C(\rho,T) \e^4\|h\|_\infty^2.\label{ener01}\end{multline}
Integrating on time between $0$ and $t$ we obtain:
\begin{multline}\frac 1 2\|f(t)\|_2^2+\e^{-2}\int_0^t ds\|\ip f(s)\|_\nu^2[\lambda-\gamma-C(\rho,T)(C\kappa^2+C_\kappa)]\le\frac 1 2\|f(0)\|_2^2\\+\int_0^t ds\frac{\e^2}{4\gamma}\|\nu^{-1/2} g(s)\|_2^2 + CC(\rho,T)\int_0^t ds\|\P f(s)\|_2^2 + C_\kappa C(\rho,T) \e^4\int_0^t ds \|h(s)\|_\infty^2.\label{ener02}\end{multline}
By using Proposition \ref{prop2} we can replace $\int_0^t ds\|\P f(s)\|_2^2$ by the right hand side of \eqref{ass} so that \eqref{ener02} becomes
\begin{multline}[\frac 1 2-CC(\rho, T)]\|f(t)\|_2^2+\e^{-2}\int_0^t ds\|\ip f(s)\|_\nu^2[\lambda-\gamma-C(\rho,T)(C\kappa^2+C_\kappa)-CC(\rho,T)]\\\le[\frac 1 2-CC(\rho,T)]\|f(0)\|_2^2+\e^2\int_0^t ds[\frac{1}{4\gamma}+CC(\rho,T)]\|\nu^{-1/2} g(s)\|_2^2  \\+ C_\kappa C(\rho,T) \e^4\int_0^t ds \|h(s)\|_\infty^2.\label{ener03}\end{multline}
Now we choose the parameters  $\kappa$, $\gamma$ and $C(\rho,T)$ is such a way that 
\begin{eqnarray*}&&\frac 1 2-CC(\rho, T)>\frac 1 4,\\
&&\lambda-\gamma-C(\rho,T)(C\kappa^2+C_\kappa)-CC(\rho,T)>\frac 1 4,\\
&&\frac{1}{4\gamma}+CC(\rho,T)<\frac 1 4,
\end{eqnarray*}
so that \eqref{ener04} becomes
\begin{multline}\|f(t)\|_2^2+\e^{-2}\int_0^{ t} ds\|\ip f(s)\|_\nu^2\\\le \e^2\int_0^{t} ds\|\nu^{-1/2} g(s)\|_2^2+C_\kappa C(\rho,T) \e^4\int_0^{ t}\|h(s)\|^2_\infty+2\|f(0)\|_2^2.\label{ener04}\end{multline}
\end{proof}

\medskip
\noindent{ \bf Proof of Proposition \ref{prop2}:} 
\begin{proof}It can be shown  along the lines of \cite{EGKM}.
We consider the following weak version of \eqref{ene}, obtained by multiplying \eqref{ene} by a smooth function $\psi$, integrating for $(x,v,s)\in \O\times\R^3_v\times[0,t]$ and integrating by parts to move all the derivatives on $\psi$:
\begin{eqnarray}&&
\int_{\O\times \R^3_v} dx dv f(x,v,t)\psi(x,v,t)-\int_{\O\times \R^3_v} dx dv f(x,v,0)\psi(x,v,0)\label{weak0}
\\&&-\int_0^tds \int_{\O\times \R^3_v} dx dv f(x,v,s)\pt_t\psi(x,v,s) 
+\frac 1{2}\int_0^tds\int_{\O\times \R^3_v} dx dv f(x,v,s)\psi(x,v,s)\pt_t\log\mu \notag
\\&&-\frac 1 \e\int_0^tds\int_{\O\times \R^3_v} dx dv f(x,v,s)v\cdot \nabla\psi(x,v,s)\notag\\&&+ \frac 1{2\e}
\int_0^tds\int_{\O\times \R^3_v} dx dv f(x,v,s)\psi(x,v,s)v\cdot \nabla\log\mu \notag
\\&&
+\frac 1 {\e^2} \int_0^tds\int_{\O\times \R^3_v} dx dv \psi(x,v,s) (L f)(x,v,s) =\int_0^tds\int_{\O\times \R^3_v} dx dv \psi(x,v,s) g(x,v,s),\notag
\end{eqnarray}
for smooth test function $\psi$. We apply this for $\psi =\sqrt{\mu}\zeta$. Then from \eqref{weak0} we get
\begin{multline}
\int_{\O\times \R^3_v} dx dv f(x,v,t)\sm \zeta(x,v,t)-\int_{\O\times \R^3_v} dx dv f(x,v,0)\sm \zeta(x,v,0)\\-\int_0^tds\int_{\O\times \R^3_v} dx dv f(x,v,s)\sm\pt_t\zeta(x,v,s) -\frac 1 \e\int_0^tds\int_{\O\times \R^3_v} dx dv f(x,v,s)\sm v\cdot \nabla\zeta(x,v,s)\\
+\frac 1 {\e^2} \int_0^tds\int_{\O\times \R^3_v} dx dv \sm \zeta(x,v,s)(L f)(x,v,s) =\int_0^tds\int_{\O\times \R^3_v} dx dv g(x,v,s) \sm \zeta(x,v,s).\label{weak}
\end{multline}
We remark that the {\it bad} term disappears due to a cancellation.
We can write \be\P f= [a + c(|v|^2-3T)]\sm,\label{Pf}\ee for some funtions $a(x,t)$, $c(x,t)$ such that
\be \int_{\O}dx a(x,t)=0, \quad \int_{\O}dx c(x,t)=0.\label{aver}\ee
The conditions \eqref{aver} are satisfied in consequence of the assumption \eqref{normal}.

To get bounds on $a$ and $c$ we use some particular functions $\zeta_a$ and $\zeta_c$. In fact we choose
\be \zeta_a= (|v|^2-\b_a) v\cdot\nabla \phi_a,\label{deltaa}\ee
\be \zeta_c= (|v|^2-\b_c) v\cdot\nabla \phi_c,\label{deltac}\ee
where $\phi_a$ solves
\be -\Delta \phi_a=a,\ee
and $\phi_c$ solves
\be -\Delta \phi_c=c,\ee
and $\b_a$ and $\b_c$ are constants to be chosen as in \cite{EGKM,EGKM1}. The zero average conditions for $a$ and $c$ \eqref{aver} are essential to ensure the solvability of \eqref{deltaa} and \eqref{deltac} and they are compatible with the equation as a consequence of the \eqref{normal}.
The estimates \eqref{ass} of $\|a\|_2$ and $\|c\|_2$ are obtained as in \cite{EGKM,EGKM1}.
\end{proof}
\subsection{The non linear problem}
Now we remind that $g$ is given by \eqref{defg}. The strategy to construct the solution to \eqref{eqf} is to define a sequence $\{f^{(n)}\}_{n=0}^{n=\infty}$ of solutions to the following {\it linear} problems: $f^{(0)}=0$ and, for $n>0$
\be\pt_t  f^{(n)} +\e^{-1} v\cdot \nabla  f^{(n)}+ \frac 1 2f^{(n)}[(\pt_t+\e^{-1} v\cdot \nabla) \log\mu] +\e^{-2} {L}  f^{(n)} =g^{(n)},\label{enen}\ee
with given $g^{(n)}\in L^2(\Omega\times \R_v)$. In view of \eqref{eqf} the choice of $g^{(n)}$ will be
\be g^{(n)}= \e^{-1}L^1 f^{(n-1)}+\e^{\frac 1 2}\G(f^{(n-1)},f^{(n-1)}) +\e^{-\frac 1 2} A,\label{defgn}\ee
for $n\ge 1$. In consequence, we also define $h^{(n)}$ so that
\be f^{(n)}=((1+|v|^2)^{ -\sigma}\sqrt{\frac{\mu_M}{\mu }}h^{(n)}
,\label{defhn}\ee
By \eqref{defgn}, we have
\be \P g^{(n)} =0. \label{Pg=0n}\ee

We have the following
\begin{lemma}\label{lmmag}
\begin{eqnarray} &\|\nu^{-1/2}\G(f,f)\|_2\le C\|f\|_2\|h\|_\infty,& \|\nu^{-1/2}\G(f,f)\|_2\le C\|h\|_\infty^2;\\
& \|L^1 f\|_2\le C(\rho,T)\|f\|_2, & \|L^1 f\|_\infty\le C(\rho,T)\|h\|_\infty;\label{estL1}\\
& \|A\|_2 \le C(\rho, T),& \|A\|_\infty\le C(\rho, T).\label{estAA}\end{eqnarray}
\end{lemma}
The proof of the lemma follows as in \cite{EGKM1}, using also Proposition \ref{macrosc} for \eqref{estL1} and \eqref{estAA} and we do not repeat it.

As a consequence, reminding \eqref{defgn}, and using $\|f^{n-1}\|_\infty\le C \|h^{n-1}\|_\infty$, we have 
\begin{multline} \e^2\int _0^tds\|g^{(n-1)}(s)\|_2^2\le \\ C(\sup_{0\le s\le t}\e^{3/2}\|h^{(n-1)}(s)\|_\infty)^2\int_0^t ds \|f^{(n-1)}(s)\|_2^2 +C(\rho,T)^2\int_0^t ds\|f^{(n-1)}(s)\|_2^2 + t C(\rho, T)^2\e,\label{gn2}\end{multline}
\be \e^{7/2}\sup_{0\le s\le t} \|\tilde g^{n-1}\|_\infty \le \e C(\e^{3/2}\sup_{0\le s\le t}\|h^{n-1}(s)\|_\infty)^2 + \e C(\rho,T)\e^{3/2}\sup_{0\le s\le t}\|h^{n-1}(s)\|_\infty+ \e^2C(\rho,t).\label{gninf}\ee
By using \eqref{gn2} and \eqref{ener040} we have
\begin{multline}\|f^{(n)}(t)\|_2^2+\e^{-2}\int_0^{ t} ds\|\ip f^{(n)}(s)\|_\nu^2\le (\sup_{0\le s\le t}\e^{3/2}\|h^{(n-1)}(s)\|_\infty)^2\int_0^t ds \|f^{(n-1)}(s)\|_2^2\\ +C(\rho,T)^2\int_0^t ds\|f^{(n-1)}(s)\|_2^2 +\e tC_\kappa C(\rho,T) (\sup_{0\le s\le t}\e^{3/2}\|h^{(n-1)}(s)\|_\infty)^2 +2\|f(0)\|_2^2+ t C(\rho, T)^2.\label{ener041}\end{multline}
By using \eqref{gninf} and \eqref{esth} we have
\begin{multline}\sup_{0\le s\le t}\|\e^{3/2} h^{(n)}(s)\|_\infty^2\le C^2\Big[\e^{3/2}\|h_0\|_\infty+\sup_{0\le s\le t}\|f^{(n)}\|_2\\ + \e C(\e^{3/2}\sup_{0\le s\le t}\|h^{(n-1)}(s)\|_\infty)^2 + \e C(\rho,T)\e^{3/2}\sup_{0\le s\le t}\|h^{(n-1)}(s)\|_\infty+ \e^2C(\rho,t)\Big]^2.\label{esth1}\end{multline}

We assume that 
\be \|f(0\|_2^2\le c_0\ll 1, \quad \e^{3/2}\|h(0)\|_\infty^2\le c_1\ll1.\ee

\noindent{\bf Inductive hypothesis}: Fixed $\bar t>0$, assume
\be \sup_{0\le \ell\le n-1}\sup_{0\le t\le \bar t}\|f^{(\ell)}(t)\|_2^2\le \eta_0\ll 1,\quad \sup_{0\le \ell\le n-1}\sup_{0\le t\le \bar t}(\e^{3/2}\|h^{(\ell)}(t)\|_\infty)^2\le \eta_1\ll 1.\ee

\medskip

By using this assumption we have
\be \sup_{0\le t\le \bar t}\|f^{(n)}(t)\|_2^2\le \eta_1\eta_0 \bar t +C(\rho,T)^2 \bar t \eta_0 +\e \bar t C_\kappa C(\rho,T)\eta_1^2+ 2 c_0 + \bar t C(\rho,T).\label{ener042}
\ee
We choose $\e$, $c_0$, $C(\rho,T)$, $C_\kappa$ so that $$2 c_0 + tC(\rho, T)<\frac{\eta_0} 3,\quad \e \bar t C_\kappa C(\rho,T)\eta_1^2<\frac{\eta_0}3,\quad \eta_1 \bar t +C(\rho,T)^2\bar t<\frac 1 3,$$ so that
\be \sup_{0\le t\le \bar t}\|f^{(n)}(t)\|_2^2<\eta_0.\ee
Then
\begin{multline}\sup_{0\le t\le \bar t}(\|\e^{3/2} h^{(n)}(t)\|_\infty)^2\le C^2 c_1+C\eta_0+ \e^2 C^2\eta_1^2 + \e C(\rho,T)\eta_1+ \e^4C(\rho,t)^2\Big].\label{esth12}\end{multline}
We choose $\e$, $c_1$, and $\eta_0$ so that $$C^2c_1+C\eta_0+ \e^4C(\rho,t)^2<\frac{\eta_1}2, \quad \e^2C^2\eta_1 +\e C(\rho,T)<\frac{\eta_1}2,$$ so that
\be \sup_{0\le t\le \bar t} \|h^{(n)}\|_\infty^2<\eta_1.\ee
Therefore the inductive hypothesis is verified up to $n$.
This shows that the sequence $\{f^{(n)}\}_{n=0}^{n=\infty}$ is uniformly bounded in $L^2$ by $\eta_0$ and in $L^\infty$ by $\eta_1$. By similar arguments one can show that $\|f^{(n)}-f^{(n-1)}\|_2 \le \theta \|f^{(n-1)}-f^{(n-2)}\|_2$ for some $\theta<1$  and hence the sequence is convergent and the limit solves uniquely \eqref{eqf}.
The proof of the positivity of $F$ is standard \cite{EGKM1}. This concludes the proof of Theorem \ref{mainth}.

\bigskip\bigskip

\noindent\textbf{Acknowledgements.}
 This work was supported in part by AFOSR [grant FA-9550-16-1- 0037]. PLG was supported also by the Spanish governement project FIS2013-43201P.

\end{document}